COVER FEATURE

# The Ultimate Display: Where Will All the Pixels Come From?

*Benjamin Watson*
Northwestern University

*David Luebke*
University of Virginia

**Could the answer be to compute fewer pixels? Renderers that break traditional framed patterns and opt for temporally adaptive sampling might be the key to printer-resolution wall displays that update hundreds of times per second.**

Imagine that the walls of your home and office are displays that show details sharper than your eye can possibly see and that they are updating those details so quickly that you couldn't notice any flicker or delay. Text displayed on these walls is as crisp as it is on the printed page. During a videoconference, you can see every flicker of emotion on a colleague's face. You can see the grass thrown up by a running back's feet or wisps of hair around your favorite actor's face. In computer games, you can see the grit on your opponent's racing car and enjoy an instant response to your every twitch of the steering wheel.

Building such a display would be extremely difficult. A wall-sized display that supports human visual acuity during close-up viewing would require about 670 megapixels, with 190 dpi. Why then do printers use resolutions that are so much higher? As the sidebar "How Much Detail Can We See?" describes, one reason is the eye's hyperacuities, and a wall display that supports these would require roughly 165 gigapixels at 3,000 dpi. This is more than 10,000 times the pixels available in today's largest displays.

Human temporal sensitivities to delay only increase requirements for the ultimate display: In addition to having extremely high spatial resolution, the display must also be refreshed about four times more quickly than current displays.

Obviously, we won't see such displays for a long time. In the meantime, we can pursue a more realistic and still extremely useful challenge: printer-resolution (600 dpi) wall displays updated at 240 Hz, requiring 7.2 gigapixels. For those still raising an eyebrow, such resolutions are not that far off. IBM's T221 display already has a resolution of 200 dpi, and Sony's Librie electronic book device uses a 170 dpi e-Ink technology. With tiled displays becoming increasingly common and inexpensive, it is easy to imagine assembling an array of 50 displays similar to the T221 to construct an acuity-supporting 455-megapixel display at 200 dpi, refreshed at 48 Hz.

The real challenge of these displays is bandwidth. At nearly 2 terapixels per second, printer-resolution bandwidths will be more than 10,000 times higher than those in today's common graphics systems. For those who can afford it, parallel graphics can solve much of this problem. For example, by spending roughly $20,000 to purchase 50 graphics processor units (GPUs) to feed your printer-resolution wall display, you can reduce bandwidth requirements to 24 gigapixels per second, only 200 times today's typical rates.

But parallel rendering is not enough. If the trends of the past 10 to 20 years continue, a 200-fold improvement in bandwidth will take more than six years to realize on graphics buses, about 10 years on GPUs, and more than 45 years on GPU-to-display wiring.[1]

Clearly, overcoming this bandwidth challenge any time soon will require more fundamental innovation. Exotic display technologies, such as holographic, autostereoscopic, and volume displays, demand even higher bandwidth than traditional displays and will only exacerbate the problem.

We offer one such innovation: adaptive frameless rendering. This approach abandons the traditional concepts of animation as a sequence of images, or *frames*, and departs from viewing images as temporally coherent grids of samples, or *pixels*. Instead, it generates samples selectively when and where they are needed. A prototype system indicates that our approach could reduce bandwidth requirements by one to two orders of magnitude.

## MORE PIXELS WITH MORE GPUS

One way to get more pixels than a single GPU can produce is to hook up more GPUs and produce the additional pixels in parallel. The simplest option is to buy a motherboard that supports multiple GPUs. Some high-end motherboards support the SLI (Nvidia) or CrossFire (ATI) multi-GPU protocols, but these allow using only two GPUs, which themselves must have dedicated hardware.

Several hardware and software products offer a more scalable solution—indeed, graphics architecture has a long history of using parallel processors to accelerate image generation. These products all face the same key challenge: finding an efficient way to distribute the rendering load to parallel graphics processors and then to assemble the results into a coherent image.

An artist or engineer creating 3D models defines rendered primitives (typically polygons) in 3D object space, but image generation involves a transformation into a view-defined 2D screen space. *Rasterization* then tiles the transformed polygons with pixels. *Depth buffering* determines which polygon is visible at each pixel by comparing the rasterized polygon's depth to the stored depth of the nearest polygon, if any, seen thus far at that pixel.

In a parallel graphics system, multiple processors perform transformation and rasterization. The rasterization processors write to specific pixels in screen space. However, the transformation from object space to screen space involves an essentially arbitrary reordering of polygons, so the system must sort the primitives into screen space somewhere along the way to assemble a coherent image.

### A rendering taxonomy

A popular taxonomy for parallel rendering classifies parallel graphics systems according to where in the rendering pipeline they perform the sort.[2] The taxonomy distinguishes three major sorting options.

*Sort-first* systems perform the sort before the 3D to 2D transformation. They allocate regions of the screen to GPUs and allocate the primitives that will fall on a given screen region to the corresponding GPU for transformation and rasterization. If a GPU transforms a primitive and finds that it now intersects another screen region (perhaps because the camera or object has moved since the last frame), a sort-first system sends the primitive to the new GPU.

Sort-first systems must deal with load-balancing issues, since many primitives can end up in a single screen region—for example, in a distant but complex object. Also, some primitives will cross region boundaries or span multiple regions; the sort-first system must replicate these across the relevant GPUs.

*Sort-last* systems postpone the sort until after rasterization. They allocate primitives to GPUs round-robin, which avoids the load-balancing issues of sort-first systems, and use a distributed form of depth buffering to composite the resulting images by choosing the closest primitive at each pixel.

Because sort-last systems must deal with all pixels that all GPUs produce, they require a custom network with extremely high bandwidth for compositing. If such a network is available, sort-last systems can enable very scalable parallel rendering.

*Sort-middle* systems break the GPU into its constituent transformation and rasterization components and assign rasterizers to screen regions. They transform primitives, establishing the appropriate screen regions, and then sort the primitives to the corresponding rasterizer.

Most parallel graphics systems, including modern GPUs, use a sort-middle system internally, but it is difficult to build a sort-middle system from commodity GPUs because they integrate the transformation and rasterization stages.

### Chromium

Some software systems enable running graphics applications over a network to a cluster of multiple computers to drive a tiled display.

Chromium[3] generalizes this approach, supporting the connection of machines in almost arbitrary ways with streams of graphics commands. For example, the system can implement sort-first rendering by intercepting a stream of graphics commands from an application using the OpenGL graphics application programming interface (API). It then routes the primitives to the machines in the cluster, each as its own stream of OpenGL commands. Each machine renders this stream on its own GPU, which in the simplest case drives a projector in the tiled display.

Chromium also lets users composite images from different computers, for example, to render onto a single nontiled display. It does this by reading back the pixels on each machine (again an OpenGL command) and sending them as yet another OpenGL stream to be collected and combined on yet another computer. By reading back depth as well as color information, this compositing process enables sort-last rendering.

Chromium has proven to be a flexible, powerful platform in practice and is now in use at universities and research labs around the world.

### Compositing solutions

For high enough resolutions, the bandwidth of shipping all those pixels around can easily saturate even gigabit networks. This is the infamous fan-in problem of parallel graphics: No matter how finely you slice the image-generation process among multiple machines, eventually you must collect all the results into a single image, and, invariably, the interconnect becomes a bottleneck.

Several hardware solutions attempt to address this problem. Both SGI's Onyx4 and Hewlett Packard's sv7 workstations combine in a sort-first fashion the output of parallel graphics pipelines that are rendering to nonoverlapping screen regions.

Other companies, such as ORAD, provide products based on the digital visual interface (DVI) digital video standard that offer a variety of sort-first configurations for commodity GPUs. The PixelFlow custom graphics supercomputer[4] introduced sort-last rendering, implemented with a linear network of rendering processors that performed custom compositing at very high speed. The prototype Lightning-2 compositing system[5] supported sort-first and sort-last rendering across commodity GPUs using DVI. The Metabuffer project[6] followed a similar concept. HP's Sepia project also aims to provide sort-last rendering, and this prototype may well become a product in the near future.

Each of these hardware solutions works with a software layer such as Chromium or various vendor-supplied libraries and toolkits such as OpenGL Multipipe SDK. Currently, there is no clean or cross-platform solution for specifying the compositing architecture (sort-first, sort-last, or a hybrid), including if and how to combine the results to drive a tiled display. There is also no such solution for supporting different types of compositors (network, DVI-based, custom, and so on).

The Parallel Image Compositing API (PICA) effort spearheaded at Lawrence Livermore National Lab seeks to address all these issues. PICA abstracts the composition process as a graph of compositing nodes specified on partial image "framelets" and provides a clean, powerful API.[7]

## FEWER PIXELS ON THE WIRE

Even if developers can lash together enough GPUs to drive a printer-resolution wall display, they face the issue of bandwidth *to* the display. In fact, developers are already encountering this issue on the commonplace DVI standard.

Apple Computer's flagship monitor, the 30-inch Cinema HD display, uses a standard DVI cable, but few graphics cards on the market can drive the display's 2,560 × 1,600 pixel resolution because it requires a card with dual-link DVI. This higher capacity version of the standard is available on only a handful of cards and is intended primarily for the workstation market. Driving a gigapixel display refreshed at 240 Hz (24 bits per pixel) would require a shocking 5.6 terabits per second.

One obvious solution is to compress the video stream before sending it over the wire and to decompress it locally within the display. Many remote and scalable rendering systems already take a step in this direction by incorporating on-the-fly compression of imagery to be sent over the network, with decompression on the local GPU driving the display. The high speed with which this must take place presents a challenge for the scale of displays we are discussing and, of course, still leaves the local GPU updating every pixel every time on its display. As pixel density and number of pixels rapidly increase, the display interface itself—literally the wires from GPU to monitor and the protocols that run over them—becomes the limiting factor.

The recent digital packet video link (DPVL) standard by the Video Electronics Standards Association seeks to address this limit (http://www.vesa.org/Standards/summary/2004_4.htm). DPVL provides a selective refresh interface in which the host GPU can transmit only modified regions of a screen. For example, an ultrahigh-resolution monitor might be used to simultaneously display both a text region that requires high spatial resolution but comparatively infrequent updates and a graphics region that requires high update rates, but perhaps at the expense of spatial resolution.

With the DPVL interface, the host can update these regions as separate packets and send only the packets that correspond to a given region or subregion when that region's image content changes. Furthermore, DPVL defines protocols for the packet metadata that the display uses to composite the packets into a final image. Examples of packet metadata might include scaling information—for example, scaling a video packet beyond its native resolution—or gamma correction tables.

Such a selective update scheme will substantially improve the usability of very high resolution monitors. If the entire display is used for a single video stream, however, such as a 3D graphics application, DPVL alone will not benefit the rendering system. Gigapixel imagery requires content-sensitive methods that can selectively update the images themselves, not just portions of the display.

## FEWER PIXELS IN THE FIRST PLACE

Rather than simply harnessing the bandwidth problem to dozens of GPUs or compressing the pixels that cause much of it, researchers should address the problem at its source and compute fewer pixels. Interactive rendering possesses high temporal coherence: Most pixels—more generally, most samples—stay relevant for several frames. Several researchers have investigated *sample-reuse* strategies that build on this observation to reduce rendering bandwidth.[8-12] Adaptive frameless rendering combines some of these strategies with innovations to produce one of the most successful low-bandwidth renderers to date.

### Sample reuse

Interactive *ray tracing* will soon make sample-reuse techniques especially relevant. The ray tracing image-synthesis technique computes color at a sample point (typically a pixel center) by shooting a geometric ray through the sample into the virtual 3D scene and determining which object the ray hits first. Unlike current GPU-based rasterizing renderers, ray tracing can sample the depicted scene selectively—a crucial capability if the tracing is to produce only samples that are truly needed. Thanks in part to clever software techniques that have improved ray tracing's memory locality, ray tracing has become interactive on successively less expensive hardware, including supercomputers, PC clusters, GPUs, custom chips, and laptop CPUs.[13]

Adaptive frameless rendering is based on the original frameless rendering approach.[8] This sample-reuse system replaced the simultaneous, double-buffered update of all pixels in a sequence of frames with the continual, single-buffered update of pixels distributed stochastically in space, each representing the most current input when the sample was taken. Because the system immediately sends these up-to-date pixels to the front buffer, frameless rendering has very low latency response to user input: The very moment the user moves the mouse, joystick, or head-mounted display, some pixels on the display will change to reflect that movement. However, as Figure 1 shows, temporal incoherence causes visual artifacts in dynamic scenes: Moving objects tend to dissolve into sparkles or noise.

*Figure 1. Prototype images showing just adaptive frameless sampling on the left, and the result of also applying adaptive reconstruction to those samples on the right. The visual quality of the resulting imagery is similar to what a framed renderer can produce, but at an order of magnitude fewer samples per second.*

Other researchers have explored different ways to loosen traditional framed spatial and temporal sampling constraints and reuse older samples. For example, just-in-time pixels[14] uses a new temporal sample for each scanline, while the address-recalculation pipeline[15] sorts objects into several layered frame buffers refreshed at different rates. The Talisman architecture[9] renders portions of the 3D scene at different rates, and composites them into a single image.

Sample reprojection reuses samples from previous frames by repositioning them to reflect the current viewpoint. Render Cache[10] reprojects and reconstructs temporally incoherent samples on the CPU, using depth comparisons and fixed 3 × 3 and 7 × 7 filtering kernels. It uses a priority image to guide new samples toward regions that have not been recently sampled, are sparsely sampled, or contain temporal color discontinuities. Tapestry[11] uses a hardware-interpolated 2.5D mesh to reconstruct samples, cached as mesh vertices, on the GPU. The system also uses a priority image to guide new samples toward spatial color and depth discontinuities. Shading Cache[12] instead stores samples as vertices in the 3D scene itself. The system directs samples toward spatial color discontinuities as well as toward specular and moving objects.

Although the images that these systems produce combine samples created at many different moments, they all differ from frameless rendering because they sample at regular intervals corresponding to each frame.

### Adaptive frameless rendering

Like the original frameless rendering,[8] our adaptive frameless rendering[16] updates the image one sample at a time, computing each update using the most current input, and abandoning the notions of pixels and frames. Freed of these rigid traditional sampling patterns, the system's adaptive components can respond with fine granularity to spatiotemporal color change, permitting extremely selective updates of interactive images.

Sampling focuses not only on where the scene changes (edges), but also on when it changes (motion). As Figure 2 shows, reconstruction emphasizes new samples where the scene is dynamic to construct up-to-date imagery, and gives old samples more significant weight ~~when~~ where the scene is static to construct sharper and eventually antialiased imagery. The resulting system not only renders more accurately than the traditional renderers using the same sampling rate depicted in Figure 2~~1~~, but it is also is as accurate as traditional renderers that use an order of magnitude more samples.

*Figure 2. Comparison of nonadaptive (top row) and adaptive frameless (bottom row) rendering using the same sampling rate. Images in the left column depict a dynamic moment in which the camera is moving rapidly. Adaptive frameless rendering emphasizes recent samples, for example, successfully extracting the glasses that, in the nonadaptive view, cannot be identified. Images in the right column depict a static scene a few moments after the camera has stopped moving. Adaptive frameless rendering produces an antialiased image by incorporating the effect of older samples, while nonadaptive rendering is restricted to a single sample per pixel.*

**Sampling.** Our adaptive frameless sampler has three primary components: a controller, deep buffer, and ray tracer. The controller observes current scene content and directs the ray tracer to sample image edges more densely and image motion more frequently. It also streams samples that the ray tracer produces to both the reconstructor and the deep buffer. The deep buffer temporarily stores samples so that the controller can locate and respond to scene edges and motion.

The controller tracks scene content using a spatial tiling of the deep buffer. Because sampling is frameless, scene content is changing constantly, and the controller must continually alter this tiling. To organize these tile changes, the controller uses a K-D tree, with a cut across the tree describing the current tiling. A given cut will contain tiles of various sizes (and various tree levels), with each tile covering an image region of (roughly) equal importance.

To alter the tiling, the controller merges two unimportant tiles and splits one important tile until tile importance is again in some equilibrium. The result is that small tiles emerge over important image regions containing edges and motion, while large tiles appear in less important regions.

The controller currently uses spatiotemporal color variation to measure importance. Image edges represent changes in color over space, causing spatial color variation in the deep buffer. Motion in an image represents a color change over time, causing temporal color variation in the deep buffer. The controller also measures spatial and temporal color gradients within each tile so that it can adjust its adaptive response, using a technique from control engineering. When the scene is particularly dynamic and temporal color gradients dominate, the spatial distribution of importance is harder to track, and the number of tiles in the current cut drops, limiting adaptive response. When the scene is static and spatial gradients dominate, it is easier to track importance, and the number of tiles increases.

To determine where to sample next, the controller picks a tile at random, and then stochastically selects a location within that tile. Because tile size declines as importance increases, the controller will sample important regions more often. At the same time, it will not ignore emerging image features because it will eventually sample even large tiles (over currently unimportant image regions). As it forms each new sample, the controller also reprojects several older samples to keep deep buffer content current.

At each display refresh, the sampler sends the reconstructor its current tiling and each tile's spatial and temporal color gradients—information the reconstructor uses to perform adaptive reconstruction.

The primary difference between our adaptive frameless sampler and the original frameless sampler[8] is adaptive response: Samples are located not stochastically, but next to scene edges and motion. Our sampler differs from reprojecting renderers like Render Cache[10] in its use of a frameless sampling pattern, which enables adaptive response to scene changes with extremely low latency.

**Interactive space-time reconstruction.** Frameless sampling strategies demand a rethinking of the traditional computer graphics image concept, since the samples used to form the current image represent many different moments in time. The original frameless work[8] simply displayed the most recent sample at every pixel, resulting in a noisy image that appeared to sparkle when displaying dynamic scenes. In contrast, our reconstructor convolves frameless samples with adaptive space-time filters to extract a coherent image. This temporally adaptive reconstruction represents a substantial difference from prior reprojecting renderers.

Our adaptive frameless reconstructor has two components: its own deep buffer and an adaptive filter bank. The deep buffer stores samples arriving from the sampler in a manner that allows efficient reconstruction on the GPU. At each display refresh, the system reprojects the deep buffer's samples according to the current view and uses the adaptive filter bank to reconstruct a coherent image from them, using the tiling and gradients the sampler provides as additional input.

Where the scene is changing rapidly, the reconstructor uses filters that emphasize new samples, resulting in up-to-date but potentially blurry imagery. Where the scene is static, filters give more weight to old samples, resulting in sharp, antialiased imagery. The adaptive filter bank accomplishes this by sizing filters according to local sampling density and then shaping them according to local color gradients. Where the scene is dynamic, temporal color gradients are high, and temporal filter support is correspondingly narrow. Where the scene is static, low temporal gradients increase temporal filter support. Similar logic adjusts spatial filter support according to spatial color gradients.

To perform reconstruction on the GPU, our reconstructor treats each deep buffer sample as a *point sprite*, or rectangle rasterized at the transformed location of a given vertex. Our system stores samples as an array of vertices, reprojects them using a GPU vertex program, and filters them by rasterizing the corresponding sprites with pixel programs that contain a "scatter" implementation of the local adaptive filter—spreading the influence of the sample across neighboring pixels.

**A prototype.** The results from comparing a software prototype of our adaptive frameless renderer to both framed and nonadaptive frameless approaches are encouraging. At the same simulated sampling rates—100,000, 400,000 and 800,000 samples per second—adaptive frameless rendering achieves three to four times the accuracy (using RMS error) of these other renderers. Even more compelling, adaptive frameless rendering is as accurate as—and sometimes more accurate than—a framed renderer that uses a sampling rate an order of magnitude higher.

The ultimate display will have enormous resolution requirements. Office walls and desks should be active display surfaces with resolution comparable to laser printers, updated hundreds of times per second. If researchers are to realize such displays, they face a pressing question: Where will all the pixels come from?

We can start by tying multiple graphics processors together and computing extremely large images in harness. When bandwidth to the display itself is a limiting factor, emerging display protocols should enable selective refresh of portions of the image. Ultimately, however, recomputing and resending every pixel every frame is simply a bad idea.

We advocate a revival of strategies that reuse the image-space samples, such as frameless rendering and Render Cache, but based on temporally adaptive sampling and reconstruction. This adaptive frameless rendering improves on framed and nonadaptive frameless rendering by focusing sampling on regions of spatial and temporal change, by applying adaptive reconstruction that emphasizes new samples when scenes are changing quickly, and by incorporating older samples when scenes are static.

We are currently studying the implications of adaptive frameless rendering on rendering and display hardware. Current graphics architectures, whether sort-first, sort-middle, or sort-last, share a common synchronization constraint: Results from all parallel rendering nodes must be combined to construct each frame. Frameless sample reuse breaks this constraint, enabling truly asynchronous parallel graphics with great potential for GPU hardware.

We are also exploring display devices that take as input streams of frameless samples instead of pixels organized into frames, performing reconstruction on the display itself rather than in the GPU. In such displays, pixels might operate as a sort of systolic array through which incoming samples diffuse and compete.

This is exactly the sort of content-sensitive, selective image update that we believe the ultimate display will require. By sending only adaptively generated frameless samples to the display, we hope to radically reduce required GPU-to-display bandwidths. ☐

**References**
1. N. Govindaraju and D. Manocha, "Efficient Visibility-Based Algorithms for Interactive Walkthrough, Shadow Generation, and Collision Detection," doctoral dissertation, Univ. of North Carolina, 2004; http://www.cs.unc.edu/~naga/thesis/main.pdf.
2. S. Molnar et al., "A Sorting Classification of Parallel Rendering," *IEEE Computer Graphics and Applications*, July 1994, pp. 23-32.
3. G. Humphreys et al., "Chromium: A Stream-Processing Framework for Interactive Rendering on Clusters," *ACM Trans. Graphics* vol. 21, no. 3, 2002, pp. 693-702.
4. S. Molnar, J. Eyles, and J. Poulton, "PixelFlow: High-Speed Rendering Using Image Composition," *Computer Graphics*, vol. 26, no. 2, 1992, pp. 231-240.
5. G. Stoll et al., "Lightning-2: A High-Performance Display Subsystem for PC Clusters," *Computer Graphics*, vol. 35, no. 2, 2001, pp. 141-148.
6. W.J. Blanke et al., "The Metabuffer: A Scalable Multiresolution Multidisplay 3D Graphics System Using Commodity Rendering Engines," tech. report TR2000-16, Univ. of Texas at Austin, 2000.
7. R. Frank, "PICA: Parallel Image Compositing API Overview," Lawrence Livermore Nat'l Lab; http://www.llnl.gov/icc/sdd/img/images/pdf/PICA.pdf.
8. G. Bishop et al., "Frameless Rendering: Double Buffering Considered Harmful," *Proc. ACM Siggraph*, ACM Press, 1994, pp. 175-176.
9. J. Torborg and J. Kajiya, "Talisman: Commodity Reality Graphics for the PC," *Proc. ACM Siggraph*, 1996, pp. 353-363.
10. B. Walter, G. Drettakis, and D.P. Greenberg, "Enhancing and Optimizing the Render Cache," *Proc. Eurographics Workshop on Rendering*, 2002, pp. 37-42.
11. M. Simmons and C. Séquin, "Tapestry: A Dynamic Mesh-Based Display Representation for Interactive Rendering," *Proc. Eurographics Workshop on Rendering*, Springer Verlag, 2000, pp. 329-340.
12. P. Tolé et al., "Interactive Global Illumination in Dynamic Scenes," *ACM Trans. Graphics*, vol. 21, no. 3, 2002, pp. 537-546.

13. P. Shirley and P. Slusallek, "Introduction to Real-Time Ray Tracing," course notes, ACM Siggraph 2005 course 38, 2005.
14. M. Olano et al., "Combatting Rendering Latency," *Proc. ACM Interactive 3D Graphics*, ACM Press, 1995, pp. 19-24.
15. M.J.P. Regan and R. Pose, "Priority Rendering with a Virtual Reality Address Recalculation Pipeline," *Proc. ACM Siggraph*, 1994, pp. 155-162.
16. A. Dayal et al., "Adaptive Frameless Rendering," *Proc. Eurographics Rendering Symp.*, 2005, pp. 265-275, June, 2005.

**Benjamin Watson,** currently an assistant professor of computer science at Northwestern University, will soon become an associate professor of computer science at North Carolina State University. His research interests include adaptive display and the intersections between graphics and perception, design, and interaction. Watson received a PhD in computer science from Georgia Institute of Technology's Graphics Visualization and Usability Center. He is co-chair of the 2006 ACM Interactive 3D Graphics and Games Conference, an associate editor of IEEE Transactions on Visualization and Graphics, a member of the ACM, and a senior member of the IEEE. Contact him at watson@northwestern.edu.

**David Luebke** is an assistant professor of computer science at the University of Virginia. His research interests are interactive computer graphics, particularly the problem of acquiring and rendering very complex real-world scenes at interactive rates. Luebke received a PhD in computer science from the University of North Carolina. He is co-chair of the Graphics Hardware 2005 conference and a member of the IEEE, ACM, and Siggraph. Contact him at luebke@cs.virginia.edu.

<begin sidebar>
## How Much Detail Can We See?

Perceptual scientists have been studying the question of how much detail we can see for centuries, and they have come up with some pretty good answers.[1]

*Visual acuity* refers to the smallest spatial detail that a person's eyes can resolve. On average, a healthy human eye can resolve details spanning one cycle per minute of visual angle. (Your thumb at arm's length spans roughly one degree of visual angle, or 60 minutes). Of course, every individual has different acuity, which is why your eye doctor makes you look at those charts.

By combining the information it gets from several overlapping receptors, the eye is sensitive to even smaller details, such as two misaligned horizontal or vertical lines. This *hyperacuity* makes the eye sensitive to certain details spanning just a few seconds of visual angle.

That is still not the end of the story—after all, in addition to three spatial dimensions, the world has a fourth *temporal* dimension.

The human eye typically loses sensitivity to flicker in a stream of images above about 60 Hz. By exploiting this limit, engineers have been able to create the illusion of temporal continuity in film, television, and computer games. But interactive display must also wrestle with human sensitivity to temporal *delay*: the time that passes between the moment of user input, and the resulting display response.

Doubtless, you've experienced extremely annoying delays above one second, such as when your PC takes too long to boot. But even 15 ms of delay slows and causes errors in interaction.[2] *Double buffering*, which is widely used to avoid the strange *tearing* artifacts that appear when only half the screen is updated, keeps each frame off screen until it is complete. But double buffering alone is enough to raise delay to 33.3 ms in 60 Hz displays. Indeed, hardcore game players often turn double buffering off to reduce delays from 33.3 ms to 16.7 ms, or with 120 Hz displays, to 8.3 ms.

The eye is an amazing biological device, and the effort to saturate its range of sensitivities should keep engineers occupied for years to come.

### References

1. R. Sekuler and R. Blake, *Perception,* 4th ed., McGraw-Hill, 2001.
2. M.J.P. Regan et al., "A Real-Time Low-Latency Hardware Light-Field Renderer," *Proc. ACM Siggraph*, ACM Press, 1999, pp. 287-290.